\documentclass{article}
\usepackage{multirow}
\usepackage{adjustbox}
\usepackage{spconf,amsmath,graphicx,color, amssymb,hyperref}
\usepackage{booktabs}
\usepackage{makecell}
\usepackage{array}
\usepackage{caption}
\usepackage{subcaption}  
\usepackage{tabularx}

\usepackage{url}

\title{Aligning Generative Speech Enhancement with Perceptual Feedback}
\name{\parbox{\textwidth}{\centering Haoyang Li$^{1}$, Nana Hou$^{2}$, Yuchen Hu$^{1}$, Jixun Yao$^{3}$, Sabato Marco Siniscalchi$^{4}$, Xuyi Zhuang$^{1}$, 
\\Deheng Ye$^{5}$, Wei Yang$^{5}$, Eng Siong Chng$^{1}$\thanks{This work was supported by Tencent and Tencent-NTU Joint Research Laboratory (CENTURY), Nanyang Technological University, Singapore.}}}
\address{$^{1}$Nanyang Technological University, Singapore\\
  $^{2}$Independent Researcher\\
  $^{3}$Northwestern Polytechnical University, China\\
  $^{4}$University of Palermo, Italy\\
  $^{5}$Tencent}

\begin{document}
\ninept

\maketitle

\begin{abstract}
Language Model (LM)-based speech enhancement (SE) has recently emerged as a promising direction, but existing approaches predominantly rely on token-level likelihood objectives that weakly reflect human perception. This mismatch limits progress, as optimizing signal accuracy does not always improve naturalness or listening comfort. We address this gap by introducing a perceptually aligned LM-based SE approach. Our method applies Direct Preference Optimization (DPO) with UTMOS, a neural MOS predictor, as a proxy for human ratings, directly steering models toward perceptually preferred outputs. This design directly connects model training to perceptual quality and is broadly applicable within LM-based SE frameworks. On the Deep Noise Suppression Challenge 2020 test sets, our approach consistently improves speech quality metrics, achieving relative gains of up to 56\%. To our knowledge, this is the first integration of perceptual feedback into LM-based SE and the first application of DPO in the SE domain, establishing a new paradigm for perceptually aligned enhancement with SE.
\end{abstract}
\section{Introduction}
\label{sec:intro}
Speech enhancement (SE) is a cornerstone technology for robust human communication and human-machine interaction, supporting applications such as hearing aids, telecommunications, and voice-driven AI systems. Existing deep neural network (DNN)-based SE approaches are broadly categorized as discriminative or generative. Discriminative methods \cite{fu2019metricGAN, li2024kan, 9023218, 9413512} minimize the distance between noisy and clean speech but often generalize poorly to unseen conditions and may introduce perceptual artifacts \cite{wang2019bridging, hou2022mismatch, 9747755}. Generative methods \cite{yao2025gense, zhang2025anyenhance} instead model the distribution of clean speech to synthesize enhanced signals, improving robustness and enabling solutions to inherently generative challenges such as packet loss concealment.  

Recent advances in language models (LMs) for image and audio generation \cite{chang2023muse, wang2023neural} have inspired LM-based generative SE frameworks \cite{wang2024selm, yang2024genhancer, li2024masksr, yao2025gense, kang2025llase}, which show promising performance. These systems are typically trained with paired data under token-level prediction objectives, reconstructing clean or high-quality speech from noisy or degraded inputs. However, such objectives emphasize signal accuracy rather than perceptual aspects like naturalness and comfort, creating a fundamental misalignment with human listening preferences. 

Research on perceptual alignment for SE remains limited. MetricGAN \cite{fu2019metricGAN} introduced adversarial training to optimize perceptual metrics, such as PESQ and STOI. Nonetheless, these metrics do not always correlate strongly with human ratings \cite{santos2014improved, kumar2025using}. More recent studies incorporated MOS predictors \cite{nayem2023attention, kumar2025using}, providing valuable insights but at the cost of added complexity and modest gains. RLHF methods, such as Proximal Policy Optimization (PPO) \cite{schulman2017proximal}, also represent a promising avenue, yet those methods typically involve complex pipelines and may face stability challenges \cite{ouyang2022training}. Direct Preference Optimization (DPO) \cite{rafailov2023direct} has recently emerged as a simpler and more stable alternative. Originally developed in NLP, DPO bypasses reward modeling and reinforcement learning, directly aligning outputs with human preferences. Its strong performance in dialogue generation and summarization suggests that DPO is a compelling candidate to bridge the gap between SE objectives and perceptual quality.  

In this work, we propose Generative Speech Enhancement with Perceptual Feedback (GSEPF), the first LM-based SE approach explicitly aligned with human auditory preferences. We built on a state-of-the-art generative SE model following \cite{yao2025gense}, and fine-tune it with DPO guided by UTMOS \cite{saeki2022utmos}, a neural MOS predictor that serves as a proxy for human ratings. This approach steers the model toward perceptually preferred outputs without the overhead of traditional RLHF pipelines. Experiments on the 2020 Deep Noise Suppression Challenge test sets \cite{reddy2020interspeech} show that GSEPF delivers consistent gains in objective metrics (DNSMOS \cite{reddy2021dnsmos}, UTMOS, NISQA \cite{mittag2021nisqa}) and subjective listening tests, achieving relative improvements of up to 56\%.  

To the best of our knowledge, this is the first study to introduce DPO into speech enhancement and the first to incorporate proxy perceptual feedback into LM-based SE. Our results establish a simple yet powerful framework for perceptually aligned enhancement, signaling a paradigm shift toward enhancement systems that optimize for human auditory preference.

\section{Generative Speech Enhancement with Perceptual Feedback}

\subsection{Two-Stage Generative SE Framework}
\label{subsec:gen-se}
We build on a state-of-the-art generative SE framework \cite{yao2025gense}, formulating enhancement as a two-stage language modeling problem that integrates both semantic and acoustic representations.  

\subsubsection{Stage 1: Noise-to-Semantic (N2S) LM.}  
Given a noisy waveform $v$, the first six layers of WavLM-Large \cite{chen2022wavlm} extract continuous latent features. A pre-trained K-means model then quantizes these frame-level features into semantic tokens, $\bar{S}=\{\bar{s}_1,\ldots,\bar{s}_F\}$, where $F$ is the number of frames, and each token is the index of one of the 1024 clusters in K-means. An autoregressive language model, denoted as the N2S LM, takes $\bar{S}$ as input and produces predicted clean semantic tokens $\hat{S} = \{\hat{s}_1, \ldots, \hat{s}_F\}$, aligned frame by frame with the noisy tokens.

\subsubsection{Stage 2: Semantic-to-Speech (S2S) LM.}  
SimCodec \cite{yao2025gense}, a neural audio codec with a single codebook, encodes $v$ into acoustic tokens $\bar{A}=\{\bar{a}_1,\ldots,\bar{a}_T\}$, where $T$ is the number of frames. A second autoregressive language model, denoted as the S2S LM, takes the concatenated token sequence $\{\bar{S}, \hat{S}, \bar{A}\}$ as context and generates the enhanced acoustic sequence $\hat{A}=\{\hat{a}_1,\ldots,\hat{a}_T\}$. The enhanced waveform $\hat{x}$ is then reconstructed using SimCodec’s decoder.  

\subsubsection{Token level training objective.}  
During training, teacher forcing is applied by replacing $\hat{S}$ and $\hat{A}$ with ground-truth $S$ and $A$ obtained from the ground-truth clean speech $x$. The S2S LM is trained with the cross-entropy loss
\begin{equation}
\mathcal{L}_{\text{CE}} = -\sum_{t=1}^T \log p\left(\hat{a}_t \mid \bar{S}, S, \bar{A}, a_{1:t-1} \right).
\label{eq:ce_loss}
\end{equation}
This two-stage formulation leverages the representational power of LMs at both semantic and acoustic levels, providing a strong backbone for our perceptual alignment framework.

\subsection{Alignment with Perceptual Feedback}
\label{subsec:alignment}

\subsubsection{Direct Preference Optimization}
\label{subsec:dpo}
While cross-entropy maximization improves likelihood, it does not guarantee perceptual quality, since human preference may diverge from token-level accuracy. DPO \cite{rafailov2023direct} directly aligns outputs with preference signals via a contrastive objective.  

Given preference pairs, where one sequence $A^+$ is favored over another $A^-$ under the same context $y$, DPO maximizes the relative preference margin:
\begin{flalign}
\mathcal{L}_{\text{DPO}} = -\mathbb{E}\!\left[\log \sigma\!\left(\beta \cdot \log \frac{\pi_\theta(A^+|y)}{\pi_{\text{ref}}(A^+|y)} - \beta \cdot \log \frac{\pi_\theta(A^-|y)}{\pi_{\text{ref}}(A^-|y)}\right)\right],
\label{eq:dpo}
\end{flalign}
where $\pi_\theta$ is the trainable target model parameterized by $\theta$, $\pi_{\text{ref}}$ is a frozen reference model to stabilize learning, $\beta>0$ controls preference sharpness, and $\sigma(\cdot)$ is the logistic sigmoid. Here, $A^\pm$ are acoustic token sequences generated by the S2S LM, and $y$ denotes the conditioning context $\{\bar{S}, S, \bar{A}\}$ from Eq.~\ref{eq:ce_loss}. 

\begin{figure}[!t]
    \includegraphics[scale=0.092]{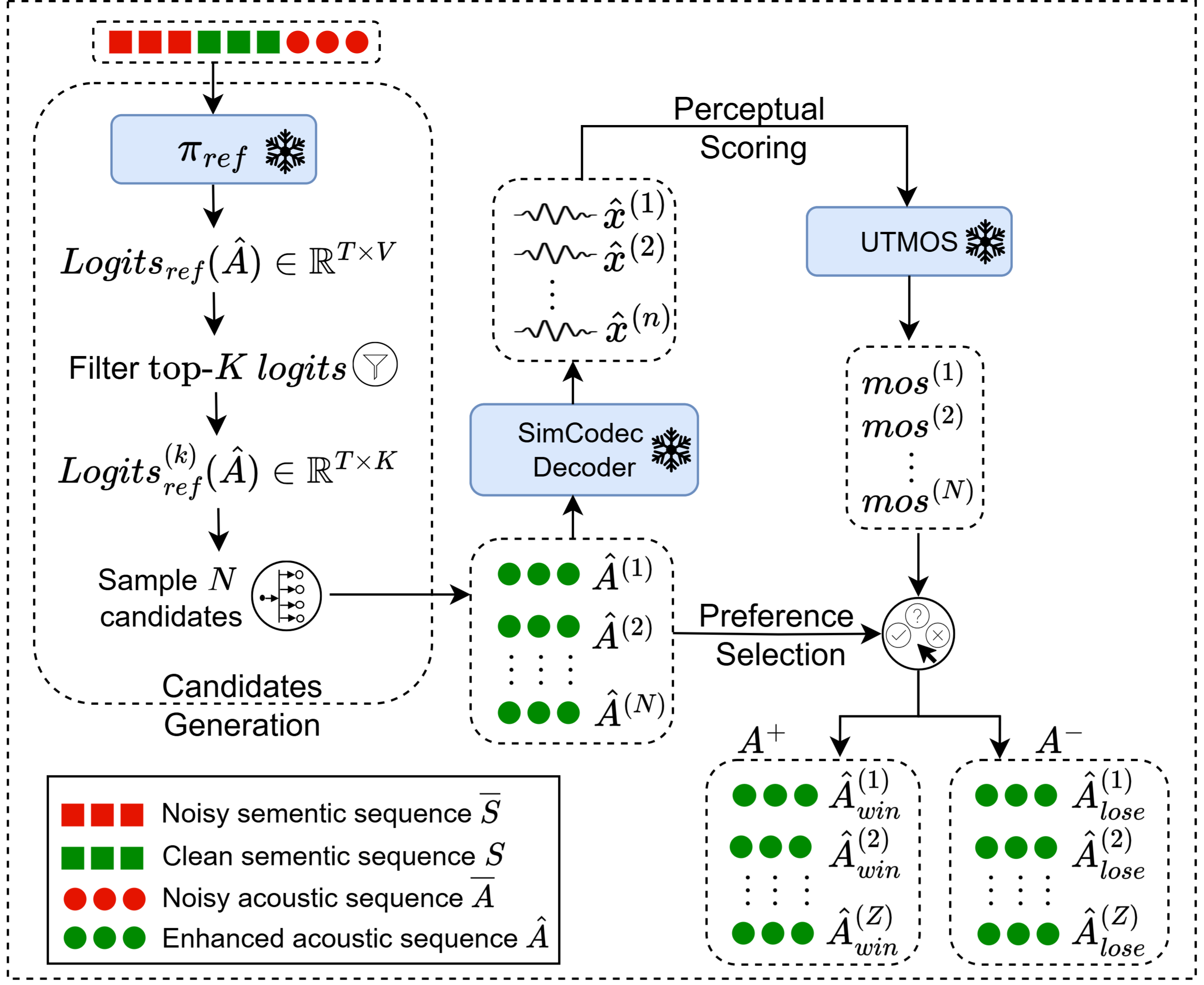}
    \caption{Pipeline to obtain preference pairs $A^+$ and $A^-$ from the reference S2S LM $\pi_\text{ref}$ during training. $A^+$ and $A^-$ are disjoint subsets of $\{\hat{A}^{(n)}\}_{n=1}^N$, each of size \( Z \), such that \( A^+ \cap A^- = \emptyset \).}
    \label{fig:sampling_pipeline} 
\end{figure}

\subsubsection{Preference Pairs Construction}
\label{subsubsec: Preference Pairs Construction}

The effectiveness of DPO depends heavily on the quality of the preferred $A^+$ and rejected $A^-$ pairs. Figure \ref{fig:sampling_pipeline} illustrates our method to obtain these preference pairs.

\textbf{Candidates Generation}. Given prompt tokens sequences $y$ (i.e. $\{\bar{S}, S, \bar{A}\}$), a pretrained S2S LM $\pi_\text{ref}$ produces the logits of the enhanced acoustic sequence $\hat{A}$ under teacher forcing following Eq \ref{eq:ce_loss}. We denote the generated logits as $Logits_{ref}(\hat{A}) \in \mathbb{R}^{T \times V}$, where $Logits_{ref}(\hat{A}) = (Logits_{ref}(\hat{a}_1), \dots, Logits_{ref}(\hat{a}_T))$ and each $Logits_{ref}(\hat{a}_t) \in \mathbb{R}^V$
denotes the logits at timestamp $t$, with $V$ the vocabulary size. Then, a top-$K$ filter is applied to $Logits_{ref}(\hat{A})$, retaining the $K$ highest-probability logits at each timestamp, producing $Logits_{ref}^{(k)}(\hat{A}) \in \mathbb{R}^{T \times K}$. From $Logits_{ref}^{(k)}(\hat{A})$, we independently sample $N$ candidate acoustic sequences. Each sequence is generated by sampling one token per timestep from the softmax-normalized probability distribution over the top $K$ logits. The resulting sequences are denoted as $\{\hat{A}^{(n)}\}_{n=1}^N$, with each sequence $\hat{A}^{(n)} = \{\hat{a}^{(n)}_1, \dots, \hat{a}^{(n)}_T\}$.

\textbf{Perceptual Scoring}. To rank the $N$ sampled sequences according to human perceptual preferences, each $\hat{A}^{(n)}$ is decoded into waveform $\hat{x}^{(n)}$ using SimCodec, then evaluated by UTMOS \cite{saeki2022utmos}, a neural MOS estimator. UTMOS provides scalable, reference-free perceptual scores $\{mos^{(n)}\}_{n=1}^N$ without the cost of extensive human evaluations, producing annotated pairs $\{(\hat{A}^{(n)}, mos^{(n)})\}_{n=1}^N$. 

\textbf{Preference Selection}. From the set $\{\hat{A}^{(n)}, \text{mos}^{(n)}\}_{n=1}^N$, the top-$Z$ sequences by MOS are designated as preferred outputs $A^+=\{\hat{A}^{(z)}_{\text{win}}\}_{z=1}^Z$, while the bottom-$Z$ form the rejected set $A^-=\{\hat{A}^{(z)}_{\text{lose}}\}_{z=1}^Z$, with $A^+\cap A^-=\emptyset$ by construction.

\subsubsection{Perceptual level training objective.}
The target S2S LM $\pi_\theta$, initialized from $\pi_{\text{ref}}$, receives the same prompt $y$ (i.e. $\{\bar{S}, S, \bar{A}\}$) to compute logits $Logits_\theta(\hat{A})\in\mathbb{R}^{T\times V}$ under teacher forcing. From $Logits_\theta(\hat{A})$ and $Logits_{ref}(\hat{A})$, sequence probabilities can be computed as
\[
p_\theta(A|y)=\prod_{t=1}^{T} p_\theta(a_t \mid a_{1:t-1},y),\quad
p_{\text{ref}}(A|y)=\prod_{t=1}^{T} p_{\text{ref}}(a_t \mid a_{1:t-1},y).
\]
Using $p_\theta(A^\pm|y)$ and $p_{\text{ref}}(A^\pm|y)$, $\mathcal{L}_{\text{DPO}}$ in Eq.~\ref{eq:dpo} is evaluated. The final training objective combines token and perceptual objectives:
\[
\mathcal{L}_{\text{overall}} = \mathcal{L}_{\text{CE}} + \mathcal{L}_{\text{DPO}}.
\]
Here $\pi_\theta$ is updated while $\pi_{\text{ref}}$ remains frozen. No scaling is applied to either loss term since their magnitudes were reasonably close.

\section{Experiments}
\subsection{Dataset}
\textbf{Trainset}. For the N2S and S2S LMs, we follow prior LM-based SE studies \cite{wang2024selm, li2024masksr, yao2025gense} by dynamically generating noisy speech using clean speech, noise clips, and room impulse responses (RIRs). Clean speech consists of a subset of LibriTTS \cite{zen2019libritts}, VCTK and the read speech partition of the 2022 DNS Challenge \cite{dubey2022icassp}, totaling $\sim$530 hours. Noise clips ($\sim$175 hours) are drawn from AudioSet and Freesound (DNS 2022), while RIRs ($\sim$ 17 hours) come from OpenSLR26 and OpenSLR28 (DNS 2022). With 40\% probability, reverberation is added using a random RIR; one noise source is mixed with 80\% probability and two sources with 20\% probability, using SNRs sampled uniformly from $[-5, 20]$ dB. SimCodec and the $K$-means tokenizer are pretrained on 960h LibriSpeech \cite{panayotov2015librispeech} and LibriTTS, respectively. All audio is resampled to 16~kHz.

\noindent\textbf{Testset}. In accordance with GenSE, we perform evaluations on the publicly available 2020 DNS Challenge \cite{reddy2020interspeech} test sets for a direct comparison with the original results.

\subsection{Evaluation Metrics}
We evaluate speech quality using DNSMOS \cite{reddy2021dnsmos}, NISQA \cite{mittag2021nisqa}, and UTMOS \cite{saeki2022utmos}. DNSMOS, a neural network-based metric, has become a standard for evaluating speech quality in LM-based speech enhancement. Unlike PESQ, which is sensitive to time misalignment \cite{wang2024selm, li2024masksr, li2025speech}, DNSMOS robustly estimates quality without a reference. Similarly to DNSMOS, NISQA and UTMOS are reference-free and correlate well with human ratings. 

To assess speaker preservation, we report speaker embedding cosine similarity (SECS), computed between ReDimNet \cite{yakovlev2024reshape} embeddings of the enhanced and ground-truth speech. 

Finally, to directly validate perceptual gains, we conduct an A/B preference test on 30 utterances from the DNS Challenge test set (w/o Reverb), where 20 volunteers compare enhanced samples before and after DPO optimization in randomized order and indicate their preference based on naturalness and listening comfort under headphone playback in quiet environments.

\subsection{Implementation Details}
The N2S and S2S LMs are decoder-only LMs consisting of 12 transformer layers, a hidden size of 1024, and 8 attention heads. The N2S LM is trained with AdamW (peak LR $1\!\times\!10^{-4}$, warmup 1k steps, cosine decay) on a single A40 GPU for 510k steps with batch size 8. The same N2S LM is used in all experiments.

For the reference S2S LM ($\pi_{\text{ref}}$), we use the same schedule but train on four A40 GPUs for 44k steps with batch size 128, stopping once DNSMOS scores saturate. This is the S2S LM used for our baseline GenSE model, denoted as GenSE\textsuperscript{*} in table \ref{tab:compare_baseline}.

Finally, the target S2S LM ($\pi_\theta$) is initialized from $\pi_{\text{ref}}$ and optimized with AdamW (LR $5\!\times\!10^{-5}$) for 400 steps with batch size 128 on a single A40 GPU, with the DPO temperature $\beta$ (Eq.~\ref{eq:dpo}) fixed at 0.1 in all experiments. Unless otherwise specified, we set $k=50$ for top-$k$ filtering, sample $N=32$ candidate sequences per prompt, and construct $Z=4$ preference pairs.

\section{Results and Discussions}

\begin{table*}[htbp]
    \caption{Comparison with baselines on DNS Challenge 2020 test sets. GSEPF achieves consistent gains across all perceptual metrics. GenSE\textsuperscript{*} denotes the reproduced performances of GenSE on our dataset. Bold = best, underline = second best.}
    \label{tab:compare_baseline}
    \centering
    \resizebox{1\textwidth}{!}{
    \begin{tabular}{@{}ccccccccccccc@{}}
    \toprule
    \multirow{3}{*}{System} &
      \multicolumn{6}{c}{w/o Reverb} &
      \multicolumn{6}{c}{w/ Reverb} \\
    \cmidrule(lr){2-7} \cmidrule(lr){8-13}
     & \multicolumn{3}{c}{DNSMOS $\uparrow$} & \multirow{2}{*}{UTMOS$\uparrow$} & \multirow{2}{*}{NISQA$\uparrow$} & \multirow{2}{*}{SECS$\uparrow$} 
     & \multicolumn{3}{c}{DNSMOS $\uparrow$} & \multirow{2}{*}{UTMOS$\uparrow$} & \multirow{2}{*}{NISQA$\uparrow$} & \multirow{2}{*}{SECS$\uparrow$} \\
    \cmidrule(lr){2-4} \cmidrule(lr){8-10}
     & SIG & BAK & OVL &  &  &  & SIG & BAK & OVL & & & \\
    \midrule
    \addlinespace[1ex]
    Noisy & 3.39 & 2.62 & 2.48 & -     & -     & -     & 1.76 & 1.50 & 1.39 & -     & -     & -     \\
    \midrule
    \addlinespace[1ex]
    GenSE \cite{yao2025gense} & 3.65 & \textbf{4.18} & \underline{3.43} & - & - & - & 3.49 & 3.73 & 3.19 & - & - & - \\
    GenSE\textsuperscript{*} & 3.65 & \underline{4.16} & 3.41 & 3.91 & 3.916 & \textbf{0.691} & 3.50 & 3.96 & 3.16 & 2.03 & 2.505 & 0.445 \\
    \midrule
    \addlinespace[1ex]
    \hspace{1em} GenSE\textsuperscript{*}\textsubscript{CE} & 3.64 & 4.15 & 3.40 & 3.91 & 3.912 & \textbf{0.691} & 3.48 & 3.96 & 3.14 & 2.10 & 2.509 & 0.452 \\
    \hspace{1em} GSEPF\textsubscript{DPO} & \underline{3.66} & \textbf{4.18} & \textbf{3.44} & \textbf{4.21} & \textbf{4.070} & 0.651 & \textbf{3.64} & \textbf{4.13} & \textbf{3.37} & \textbf{3.18} & \textbf{2.984} & \underline{0.454} \\
    \hspace{1em} GSEPF\textsubscript{CE+DPO} & \textbf{3.67} & \textbf{4.18} & \textbf{3.44} & \underline{4.17} & \underline{4.021} & \underline{0.667} & \underline{3.60} & \underline{4.10} & \underline{3.32} & \underline{2.86} & \underline{2.815} & \textbf{0.477} \\
    \bottomrule
    \end{tabular}
    }
\end{table*}

\begin{table*}[htbp]
    \caption{Performance comparison of different Preference Pairs Construction Strategies.}
    \label{tab:sampling_ablation}
    \centering
    \resizebox{1\textwidth}{!}{
    \begin{tabular}{@{}cccccccccccccc@{}}
    \toprule
    \multirow{3}{*}{System} &
      \multicolumn{6}{c}{w/o Reverb} &
      \multicolumn{6}{c}{w/ Reverb} \\
    \cmidrule(lr){2-7} \cmidrule(lr){8-13}
     & \multicolumn{3}{c}{DNSMOS $\uparrow$} & \multirow{2}{*}{UTMOS$\uparrow$} & \multirow{2}{*}{NISQA$\uparrow$} & \multirow{2}{*}{SECS$\uparrow$} 
     & \multicolumn{3}{c}{DNSMOS $\uparrow$} & \multirow{2}{*}{UTMOS$\uparrow$} & \multirow{2}{*}{NISQA$\uparrow$} & \multirow{2}{*}{SECS$\uparrow$} \\
    \cmidrule(lr){2-4} \cmidrule(lr){8-10}
     & SIG & BAK & OVL &  &  &  & SIG & BAK & OVL & & & \\
    \midrule
    \addlinespace[1ex]
    GenSE\textsuperscript{*} & 3.65 & 4.16 & 3.41 & 3.91 & 3.916 & \textbf{0.691} & 3.50 & 3.96 & 3.16 & 2.03 & 2.505 & 0.445 \\
    \midrule
    \addlinespace[1ex]
    Z=1 (Ground-truth) & 3.65 & 4.17 & 3.42 & 3.92 & 3.913 & 0.688 & 3.48 & 3.95 & 3.14 & 2.06 & 2.498 & 0.456 \\
    Z=1 & \textbf{3.67} & 4.17 & \textbf{3.44} & \textbf{4.17} & \textbf{4.052} & 0.666 & \textbf{3.62} & \textbf{4.11} & \textbf{3.35} & \textbf{2.95} & \textbf{2.873} & 0.469 \\ 
    Z=4 & \textbf{3.67} & \textbf{4.18} & \textbf{3.44} & \textbf{4.17} & 4.021 & 0.667 & 3.60 & 4.10 & 3.32 & 2.86 & 2.815 & \textbf{0.477} \\
    \bottomrule
    \end{tabular}
    }
\end{table*}

\subsection{Main Comparison with Baselines}
We explored three loss combination schemes to train the target S2S LM $\pi_\theta$: \textbf{1. GenSE\textsuperscript{*}\textsubscript{CE}.} We continue to use the cross entropy (CE) loss in Eq \ref{eq:ce_loss}, which is the original training objective of GenSE. \textbf{2. GSEPF\textsubscript{DPO}.} We use the DPO training objective in Eq \ref{eq:dpo}. \textbf{3. GSEPF\textsubscript{CE+DPO}.} We used both the CE and DPO losses. The results in Table \ref{tab:compare_baseline} show that DPO loss consistently improves all speech quality metrics, with up to 56\% increase in UTMOS (from 2.03 to 3.18) and 19\% increase in NISQA (from 2.505 to 2.984) on the w/ Reverb partition, demonstrating its effectiveness in enhancing perceptual quality rather than overfitting to UTMOS only. In addition, we observe that the DPO optimization achieves better improvement on the w/ Reverb partition than the w/o Reverb partition, which is likely because recovering clean speech from reverberant+noisy speech is more challenging and hence leaving more room for improvement. In contrast to DPO-related optimizations, we observe from the GenSE\textsuperscript{*}\textsubscript{CE} experiment that using CE loss alone does not appear to improve the target S2S LM $\pi_\theta$.  

For speaker similarity (SECS), DPO optimization results in degradation on the w/o Reverb partition but improves on the w/ Reverb partition. Combining CE loss with DPO loss improves SECS across both test partitions over only DPO loss. These results suggest that incorporating CE loss provides an anchoring effect during DPO training, discouraging over-optimization toward speech quality that compromises speaker similarity, while slightly tempering the speech quality improvements achieved by the DPO loss.

\subsection{Subjective Evaluation}
\begin{figure}[!t]
    \includegraphics[scale=0.35]{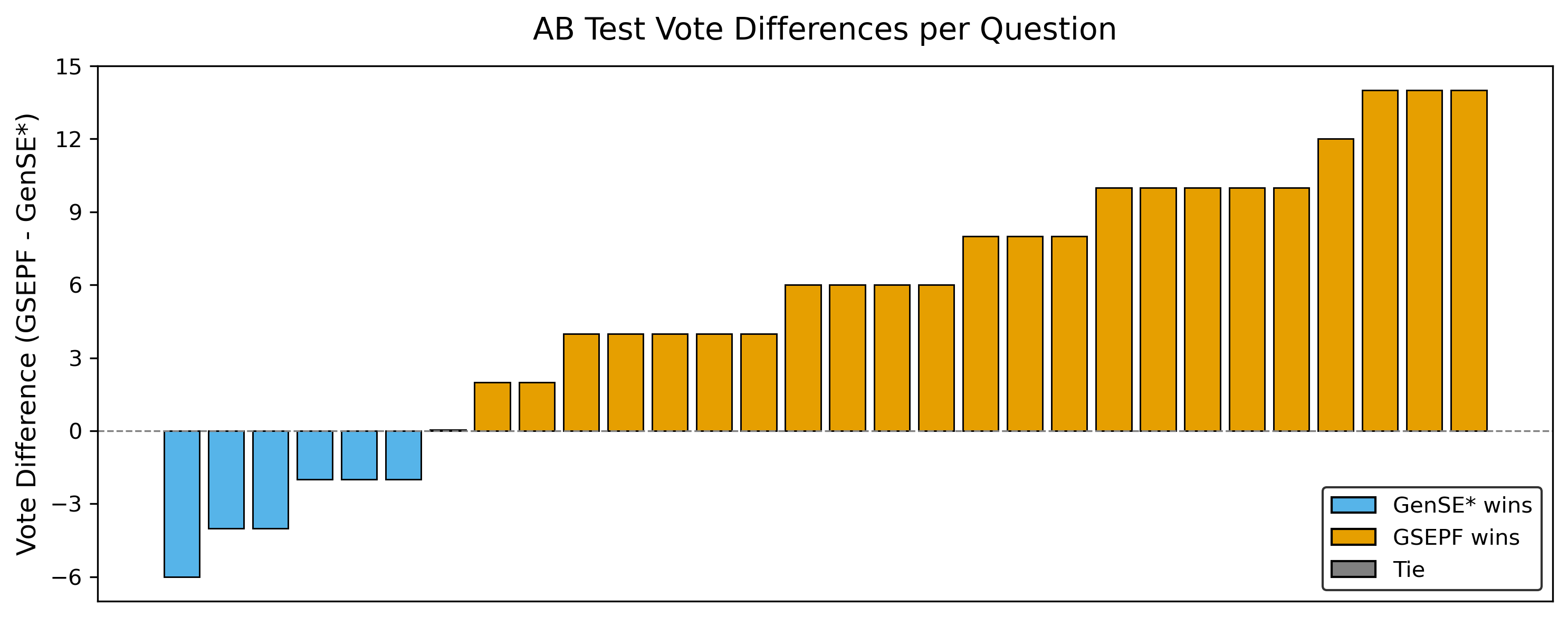}
    \caption{A/B test on naturalness and listening comfort between GenSE\textsuperscript{*} and GSEPF\textsubscript{CE+DPO}. The proposed method received 378 votes vs. 222 for the baseline, winning 23/30 cases.}
    \label{fig:ab_vote_differences} 
\end{figure}

To further validate perceptual alignment, we conducted A/B preference tests with 20 human listeners on 30 utterances. As shown in Fig. \ref{fig:ab_vote_differences}, listeners generally preferred GSEPF-enhanced samples (GSEPF\textsubscript{CE+DPO}) over the GenSE baseline ( GenSE\textsuperscript{*}) in 23 of 30 samples, citing greater naturalness and reduced listening fatigue. This strong subjective preference confirms that our DPO-based optimization aligns well with human auditory perception, beyond what can be captured by automated metrics alone.

\begin{figure}[!t]
    \centering
    \includegraphics[scale=0.47]{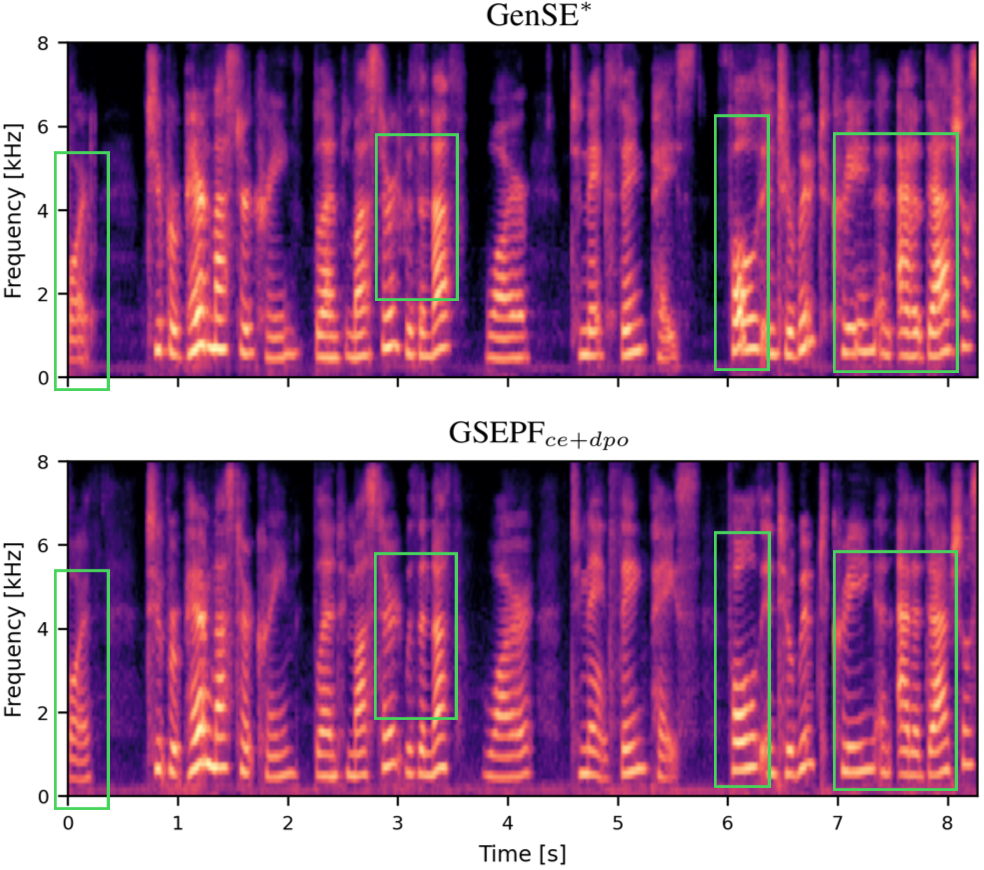}
    \caption{Example case study: GenSE\textsuperscript{*} vs. proposed GSEPF\textsubscript{CE+DPO} (mel-spectrograms). The proposed method reduces artifacts and better preserves speech harmonics compared to the baseline.}
    \label{fig:case_study} 
\end{figure}

Figure~\ref{fig:case_study} additionally presents spectrogram comparisons for an utterance with notable noise and reverberation. The GenSE baseline reduces noise but introduces artifacts in voiced regions, whereas GSEPF better preserves harmonic structures. This qualitative evidence highlights the strength of perceptual alignment: By optimizing towards human-preferred outputs, GSEPF produces speech that is both cleaner and more natural.

\subsection{Ablation Experiment: Preference Pair Selection}

In Table \ref{tab:sampling_ablation}, we conduct ablation studies on how different preference pair construction strategies impact the training of the target S2S LM $\pi_\theta$. For all experiments, we optimize using both CE and DPO loss. In the $Z$=1 (Ground-truth) experiment, we directly use the ground-truth acoustic tokens of the target clean speech $x$ as A+ instead of sampling from the reference S2S LM $\pi_\text{ref}$. We compare this setup with the $Z$=1 experiment, where the sampling of A+ follows the pipeline described in section \ref{subsubsec: Preference Pairs Construction}, with $Z$, the number of preference pairs set to 1. The results show that DPO training is ineffective when using the $Z$=1 (Ground-truth) setup. We attribute this to the model being pushed in the same direction as cross-entropy loss, thereby diminishing the unique contribution of DPO. We also investigated whether the number of preference pairs for A+ and A- have on performance, by altering the value of $Z$. From the $Z$=1 and $Z$=4 experiments, we found that while both setups effectively improved speech quality, using more preference pairs does not lead to better speech quality.

\section{Conclusion}
We presented \textbf{GSEPF}, the first framework to apply Direct Preference Optimization (DPO) to speech enhancement. By leveraging UTMOS as a proxy for human judgments and constructing preference pairs from a reference LM, GSEPF directly aligns model outputs with perceptual feedback. Despite its simplicity, it achieves up to 56\% improvement on UTMOS and clear gains on unseen metrics such as NISQA, all within only 400 training steps. This work marks a paradigm shift for LM-based SE: moving beyond token-level likelihood toward preference-driven optimization that better reflects human listening experience. Future directions include extending preference alignment to speaker similarity, controllability, and ultimately multi-objective alignment in audio and multimodal generation. 

\vfill\pagebreak
\clearpage

\footnotesize
\bibliographystyle{IEEEbib}
\bibliography{refs}

\end{document}